%% file: main.tex
\documentclass[twocolumn]{article}

\usepackage[margin=1in]{geometry}
\usepackage{times}
\usepackage{setspace}
\usepackage{graphicx} 
\usepackage{float}
\usepackage{hyperref}
\usepackage{booktabs}
\usepackage{multirow}
\usepackage{adjustbox}
\usepackage{listings}
\usepackage{xcolor}
\usepackage{multicol}
\usepackage{subcaption}
\usepackage{amsmath}

\title{\textbf{An Adaptive Multi-Agent Bitcoin Trading System}} 
\author{Aadi Singhi \\ UCL Computer Science}
\date{} 

\begin{document}
\raggedbottom
\maketitle

\begin{abstract} 
\footnote{The original thesis was submitted on 16 May 2025. This paper presents a condensed version of the full 80-page dissertation.}
This paper presents a Multi Agent Bitcoin Trading system that utilizes Large Language Models (LLMs) for alpha generation and portfolio management in the cryptocurrencies market. Unlike equities, cryptocurrencies exhibit extreme volatility and are heavily influenced by rapidly shifting market sentiments and regulatory announcements, making them difficult to model using static regression models or neural networks trained solely on historical data.
The proposed framework overcomes this by structuring LLMs into specialised agents for technical analysis, sentiment evaluation, decision-making, and performance reflection. The agents improve over time via a novel verbal feedback mechanism where a Reflect agent provides daily and weekly natural-language critiques of trading decisions. These textual evaluations are then injected into future prompts of the agents, allowing them to adjust allocation logic without weight updates or finetuning. Back-testing on Bitcoin price data from July 2024 to April 2025 shows consistent outperformance across market regimes: the Quantitative agent delivered over 30\% higher returns in bullish phases and 15\% overall gains versus buy-and-hold, while the sentiment-driven agent turned sideways markets from a small loss into a gain of over 100\%. Adding weekly feedback further improved total performance by 31\% and reduced bearish losses by 10\%. The results demonstrate that verbal feedback represents a new, scalable, and low-cost approach of tuning LLMs for financial goals.
\end{abstract}

\section{Introduction}

This paper introduces an Agentic AI system for managing a Bitcoin-only portfolio powered by DeepSeek-r1 \cite{qin2025}. Conventional machine learning methods, such as regression models and neural networks, perform well in equity markets where recurring historical patterns are linked to fundamentals such as earnings, cash flows, and balance sheet strength ~\cite{atsalakis2019bitcoin}. In contrast, Bitcoin's price is largely driven by non-fundamental forces such as regulatory announcements, influencer statements, and speculative behavior by investors driven by fear and greed. For example, in early January 2021 Bitcoin rose from about \$29,000 to over \$40,000 in less than a week following positive commentary and Tesla’s \$1.5 billion purchase, while in May 2021 it fell nearly 30\% within a week after Tesla suspended Bitcoin payments and Chinese regulators reiterated restrictions. Therefore, Bitcoin is characterized by asymmetric returns, fat tails, and high noise-to-signal ratios.\newline 

In such conditions, being frequently correct does not necessarily translate into making money. Theoretically, if a model consistently predicts the correct direction of returns, profitability should follow. However, Accuracy treats all predictions equally, whereas profitability depends on the magnitude and timing of those predictions. A model that is "often right" can still underperform if it misses major rallies or fails to exit before sharp drawdowns, while another with lower accuracy can outperform by simply capturing large moves. In such conditions, profit depends more on capital allocation, position sizing, and risk management than on raw directional correctness. Moreover, even highly accurate models may fail if execution costs, slippage, or regime shifts negate expected returns.\newline 

What is needed is a mechanism that can reason over unstructured, real-time signals rather than only fit historical patterns. LLMs offer a potential solution to this unpredictability, as they are particularly suited for large-scale textual analysis and pattern recognition. By continuously updating their understanding as new data emerges, they are better able to capture sudden shifts in cryptocurrency prices than models trained solely on historical data.\newline 

This promise, however, hinges on how we enable LLMs to evolve without full retraining. Since LLMs do not inherently update their internal knowledge, the key question is how we can refresh or refine their understanding as markets change. Existing research explores methods such as Retrieval-Augmented Generation (RAG) \cite{lewis2020rag}, Supervised Fine-Tuning (SFT) \cite{openai2023sft}, Reinforcement Learning (RL) \cite{ouyang2022instructgpt}, and Multi-Head Reinforcement Learning (MHRL) \cite{zhang2019merl}.
, but these approaches often suffer from information loss when complex human feedback is compressed into a single scalar reward signal. Several studies, including InstructGPT \cite{ouyang2022instructgpt} and Constitutional AI \cite{bai2022constitutionalai}, note that converting rich human feedback into a single scalar reward often omits contextual nuance and can cause models to optimize for the reward itself rather than the true intent of the feedback.
 To address this limitation, recent studies propose using verbal feedback: natural-language critiques provided directly to the model during inference, allowing it to retain the full semantic richness of the feedback. For instance, the Reflexion framework enables models to store and reuse their own self-critiques without any parameter updates, leading to more consistent reasoning and improved task success rates across multiple benchmarks \cite{shinn2023reflexion}. Similarly, Self-Refine introduces an iterative “generate → critique → revise” process, where the model learns from its own feedback to progressively enhance output quality \cite{madaan2023selfrefine}.
Both studies report substantial performance gains over standard LLM baselines without additional training or fine-tuning. This is especially relevant for crypto trading, where context and justification matter as much as raw predictions.\newline 

Now that this system incorporates natural-language critiques as its optimisation method, the next step is to examine how these capabilities translate into actual trading performance. Improving a trading strategy depends on identifying which input factors matter most for Bitcoin. Is it technical indicators, sentiment measures, or broader market variables? The existing literature remains inconclusive on this point, with studies reaching mixed conclusions about which features provide consistently useful signals.\newline

To build on these capabilities, the framework structures LLMs into specialised agents for technical analysis, sentiment assessment, portfolio decision-making, and strategy evaluation. Adaptability is introduced through a novel dual verbal feedback mechanism in which the Reflect agent produces natural-language evaluations of each agent’s performance at daily and weekly intervals. The trading decisions are evaluated against benchmarks such as a static baseline buy-and-hold portfolio, risk-adjusted measures such as the Sharpe ratio, and profitability indicators such as cumulative returns, regret, and accuracy. These evaluations are then injected into the next day’s or week’s prompts. Rather than adjusting model parameters or retraining, the agents incorporate this feedback as additional context in their reasoning, allowing them to refine indicator prioritisation, sentiment weighting, and allocation choices in subsequent decisions. Each agent manages its own portfolio based on distinct inputs, such as quantitative indicators or sentiment data, allowing a direct comparison of which factors lead to stronger trading performance. The system is evaluated not on classification accuracy alone but on its ability to deliver consistent profitability over time, aligning its objectives more closely with real-world investment outcomes.

\section{Background Research}
\noindent The background review is organized into two subsections: an in-depth analysis of Bitcoin, followed by a brief summary of the current research on LLMs in finance.

\subsection{Bitcoin}

Bitcoin is a decentralized digital currency introduced in 2009 by the pseudonymous Satoshi Nakamoto \cite{nakamoto2008bitcoin}. While Bitcoin is often described as a currency, it is more fundamentally an application of distributed ledger technology (DLT), where transactions are recorded across a decentralized network rather than by a central authority. Within this family of technologies, blockchain represents a specific implementation: a sequential, cryptographically linked chain of blocks that ensures immutability and transparency. Bitcoin was the first large-scale use case of blockchain. Beyond peer-to-peer payments, Bitcoin’s most profound contribution lies in demonstrating that trust can be engineered through code rather than institutions. By solving the double-spending problem via a decentralized consensus mechanism, it eliminated the need for intermediaries in value transfer \cite{narayanan2016bitcoin}. Its open, immutable, and censorship-resistant ledger redefined the notion of digital property rights, allowing individuals to hold and exchange scarce digital assets without reliance on central authorities. Moreover, its rule-based monetary policy and energy-anchored security introduced a new paradigm of programmable, verifiable money that inspired subsequent innovations in blockchain-based finance, from DeFi to asset tokenization.\newline

Bitcoin has grown into the world’s most valuable cryptocurrency (Figure~\ref{fig:bitcoin_price}), breaching \$115{,}000 in 2025 and attaining a market capitalization of roughly \$2.3 trillion with daily trading volumes exceeding \$61 billion. Its value is sustained by a fixed supply of 21 million coins, a feature often compared to gold, and it has become widely adopted both by retail participants and large institutions. Recent milestones including the approval of spot Bitcoin ETFs and its use as a corporate treasury asset have cemented its role as the benchmark of the digital asset market \cite{sec2024etf}.

\subsubsection{Price}

\begin{figure*}[t]
\centering

\begin{minipage}[t]{0.48\textwidth}
    \centering
    \includegraphics[width=\linewidth]{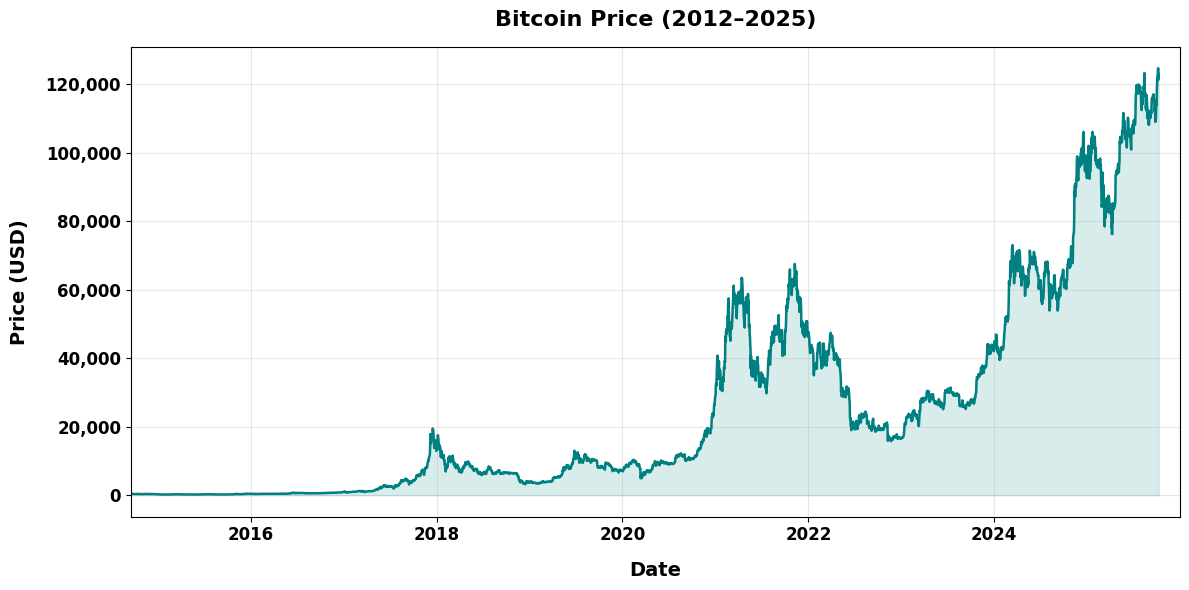}
    \caption{Bitcoin price chart}
    \label{fig:bitcoin_price}
\end{minipage}
\hfill
\begin{minipage}[t]{0.48\textwidth}
    \centering
    \includegraphics[width=\linewidth]{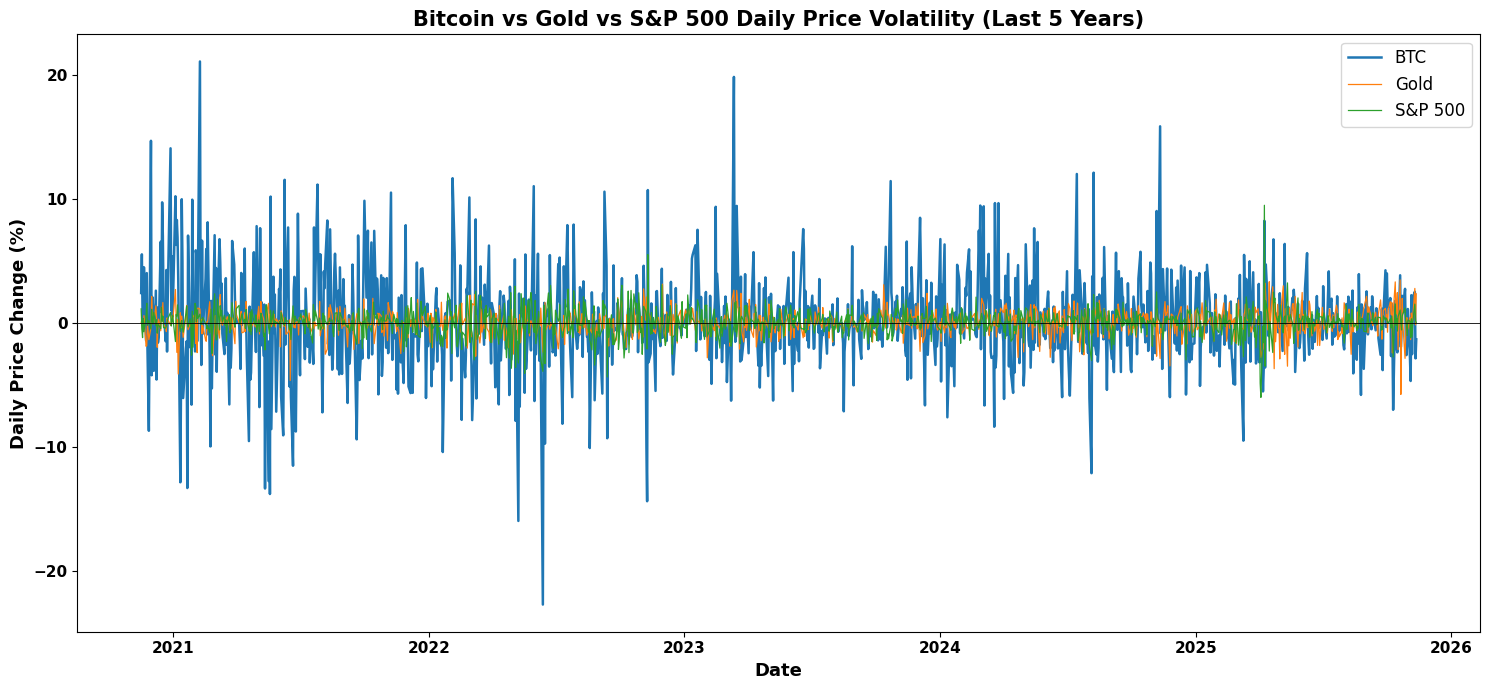}
    \caption{Historical Price Volatility (2020–2025) for Gold, S\&P 500, and Bitcoin}
    \label{fig:volatility}
\end{minipage}

\end{figure*}

\begin{table*}[t]
    \centering
    \caption{CAGR Comparison of Major Assets}
    \begin{tabular}{lccc}
        \hline
        \textbf{Asset} & \textbf{10-Year Return (\%)} & \textbf{5-Year Return (\%)} & \textbf{Sharpe Ratio} \\
        \hline
        Gold     & 10  & 7   & 0.90 \\
        S\&P 500 & 13  & 11  & 0.65 \\
        Bitcoin  & 124 & 155 & 0.96 \\
        \hline
    \end{tabular}
    \label{tab:cagr}
\end{table*}

 Over the last decade, Bitcoin has outperformed traditional assets by an extraordinary margin, growing roughly 26,900\% versus $\sim$193\% for the S\&P 500 and $\sim$126\% for gold, with a 10-year CAGR for Bitcoin of 124\% compared to 10--13\% for equities and gold as seen in Table ~\ref{tab:cagr}. \newline



But the CAGR doesn't explain the stark contrast in volatility as shown in Figure ~\ref{fig:volatility}. Bitcoin’s price movements are far more unpredictable and prone to sharp swings, making it a substantially higher risk asset relative to the traditional safe haven status of Gold and the diversified equity exposure of the S\&P 500. From an investor’s standpoint, Bitcoin rewards those with a high risk tolerance and a long horizon, while Gold is best suited for capital preservation and cushioning drawdowns. The S\&P 500 remains the core growth anchor for most portfolios. When combined, these three assets can complement one another Bitcoin driving upside potential, Gold adding stability, and the S\&P 500 delivering steady growth resulting in a diversified mix that manages risk while seeking attractive returns. Ultimately, the right balance depends on an investor’s risk appetite and time horizon.
    
\subsubsection{Dominancy}

\begin{figure}[t]
    \centering
    \includegraphics[width=\linewidth]{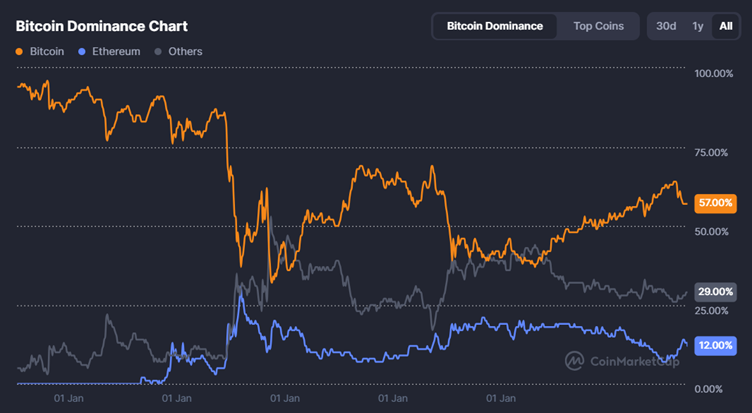}
    \caption{Bitcoin dominance as a share of total cryptocurrency market capitalization.}
    \label{fig:btc_dominance}
\end{figure}

\begin{figure}[t]
    \centering
    \includegraphics[width=1\linewidth]{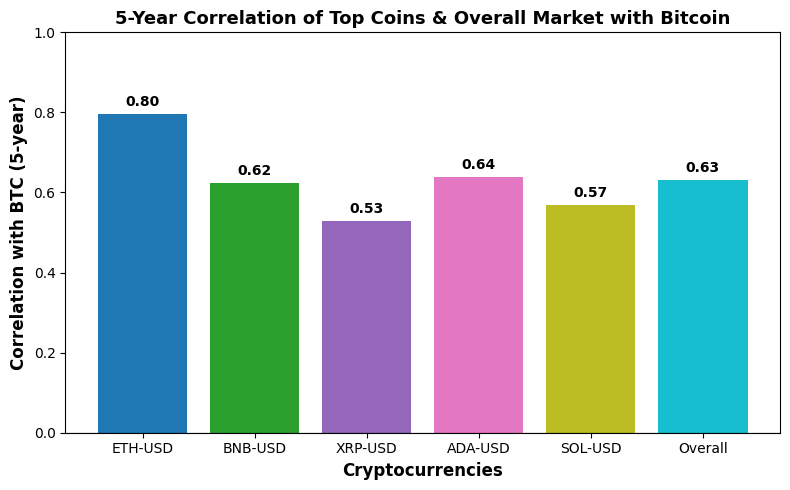}
    \caption{Average Correlations of different cryptocurrencies with Bitcoin over 5 years}
    \label{fig:correlations}
\end{figure}

Bitcoin has remained the dominant force in the cryptocurrency ecosystem since its inception in 2009, consistently accounting for 40--50\% of total market capitalization over the past decade (Figure~\ref{fig:btc_dominance}). Unlike the majority of altcoins, many of which have low survivability, speculative use cases, or heavy dependence on external platforms, Bitcoin has endured multiple market cycles without protocol failure. Moreover, empirical evidence, as seen in  (Figure~\ref{fig:correlations}), shows that Bitcoin is highly correlated with other cryptocurrencies (0.5--0.8), meaning altcoins tend to follow its price movements, particularly during downturns. The Overall market mostly moves in tandem with Bitcoin with a correlation of 0.63. This systemic co-movement undermines the rationale for intra-crypto diversification, since portfolios spread across tokens still remain overwhelmingly exposed to Bitcoin’s market trajectory. For these reasons, this paper exclusively focuses on Bitcoin, not only because of its dominance but also because it represents the most rational and risk-conscious foundation for building a profitable trading system. 

\subsubsection{Price Factors}
Bitcoin's price is mainly shaped by how limited its supply is, how investors behave, and how active the network is. The total number of bitcoins is capped at 21 million, and the reward for mining new coins is cut in half every few years. This built-in scarcity means that when more people want to buy or use bitcoin, its price tends to rise quickly. According to \href{https://www.investopedia.com/tech/what-determines-value-1-bitcoin/}{Investopedia}, bitcoin's value is driven by its supply, demand, competition from other cryptocurrencies, and investor sentiment.  

In the short term, prices move mostly because of changes in mood and expectations. News about regulation, economic trends, or social media discussions can quickly shift how investors feel, leading to sharp gains or drops. These swings often feed on themselves : optimism pushes prices higher, while fear causes sudden sell-offs \href{https://www.investopedia.com/articles/investing/052014/why-bitcoins-value-so-volatile.asp}{(Investopedia)}. 

Recent research also shows that technical indicators remain useful for understanding short-term movements in bitcoin. Empirical studies find that price and volume-based indicators often contain statistically significant predictive information in cryptocurrency markets. \cite{goutte2023deep} shows that incorporating standard technical indicators into deep-learning models improves the accuracy of crypto price forecasts relative to models using raw prices alone. Similarly, \cite{kanat2023validity} finds that classical indicators, such as moving averages and stochastic oscillators, enhance prediction performance across multiple cryptocurrencies when evaluated with machine-learning methods. More recently, \cite{jin2024technical} demonstrates that a range of technical trading rules can outperform a simple buy-and-hold benchmark under certain market conditions, indicating that technical signals still capture meaningful short-term structure in crypto price dynamics. 

Network data also plays an important role. Information such as how many transactions are happening, how many users are active, and how miners behave shows how much the system is being used in practice. Combining this with market and sentiment data gives a clearer picture of what drives bitcoin's price and whether the movements are based on real growth or short-term speculation. Together, these findings suggest that well-designed technical indicators, market sentiment, and on chain variables continue to provide practical insight into near-term bitcoin price behaviour.

\subsection{Large Language Models in Finance}

Large Language Models (LLMs) represent a major advance in natural language processing, built primarily on the Transformer architecture introduced by Vaswani et al. (2017) \cite{vaswani2017attention}. Their defining capability lies in modelling context through self-attention, which allows them to generate coherent text and perform diverse language tasks without task-specific training. As Lee et al. (2024) note in their survey of financial LLMs, this generalization capacity makes LLMs particularly powerful when dealing with unstructured, text-heavy domains \cite{lee2024finllms}. Domain-specific adaptations highlight this potential: BloombergGPT (Wu et al., 2023) trained a 50-billion-parameter model on a large financial corpus, demonstrating significant gains in sentiment analysis, question answering, and named entity recognition compared to general LLMs \cite{wu2023bloomberggpt}. Similarly, PIXIU (Chen et al., 2023) introduced FinMA, an instruction-tuned variant of LLaMA, together with a benchmark covering forecasting, sentiment, and reasoning tasks, providing evidence that finance-specific instruction tuning improves accuracy and robustness \cite{chen2023pixiu}. Collectively, these efforts show how LLMs can move beyond generic language modeling to capture the nuances of financial text, laying the foundation for their integration into trading and portfolio management systems.\newline

However, The use of LLMs in finance is not without risk. Three persistent limitations constrain their deployment in trading contexts. First, they are prone to hallucination: generating plausible but false information. In one benchmark, a finance-focused LLM produced fabricated content in 41\% of probing test cases, raising concerns about reliability \cite{lee2024finllms}. Second, their knowledge is temporally bounded by training cutoffs, making them poorly suited to tasks requiring real-time awareness unless supplemented by retrieval or live feeds. Third, they lack interpretability. With billions of parameters, it is often unclear why a particular recommendation is produced, creating difficulties for accountability in financial decision-making. Together, these weaknesses necessitate the design of auxiliary mechanisms to ensure robustness, adaptability, and transparency.\newline

Luo et al. (2025) propose a fine-tuned multi-agent system for automated crypto portfolio management, where agents trained on distinct modalities like technical signals, news, market factors, and risk metrics collaborate to build diversified portfolios \cite{luo2025llmmas}. Their architecture, which integrates expert fine-tuning with inter-agent communication, consistently outperforms both unfine-tuned GPT-4o baselines and traditional benchmarks, achieving superior Sharpe ratios and robustness. The authors argue that decomposing investment tasks into specialized, fine-tuned experts provides scalability and explainability, though at the cost of flexibility in rapidly changing conditions.

A balanced view therefore recognises both the capabilities and constraints of LLMs in financial settings. Their strength lies in processing heterogeneous information and adapting to new tasks without retraining, yet these advantages come with risks such as hallucination, improper instruction following, and lack of interpretability. Effective deployment requires careful prompt design, structured interaction protocols, and mechanisms that constrain model behaviour.

\begin{figure*}[t]
    \centering
    \includegraphics[width=1.0\linewidth]{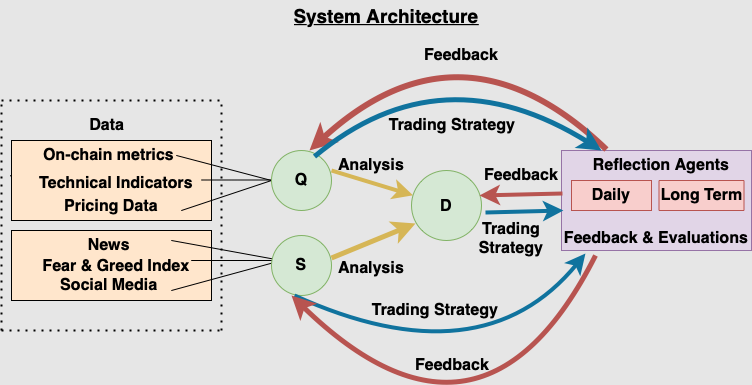}
    \caption{System Architecture}
    \label{system_arch}
\end{figure*}

\section{System Design and Methodology}

Building on the findings of the background research, the proposed system is designed not only to evaluate profitability in a portfolio across different market regimes but also to identify the contribution of individual market factors in generating profitable trades. The system architecture can be seen in figure ~\ref{system_arch}. The system prompts for each agent can be found in the appendix.

\subsection{Quantitative Agent (Quants)}
The Quants Agent is designed to analyze Bitcoin market data using quantitative methods, providing an objective assessment of market conditions. Its primary role is to process historical and real-time data, which includes price trends, technical indicators, and on-chain metrics, to classify the subsequent day’s market state as bullish, bearish, or neutral. Additionally, it generates a trading strategy (BTC/cash portfolio allocation) based on its own analysis. This strategy is derived purely from technical insights, which can help evaluate whether relying purely on technical indicators can lead to successful trade generations. This recommendation is provided to the reflection agents for evaluation, while the market classification and reasoning are forwarded to the decision agent.\newline 

It relies on structured, market-derived data to ground its predictions. This includes OHLCV data (open, high, low, close, and volume) drawn from exchange feeds ~\cite{binance2025}, as well as a range of technical indicators computed from this data using the Python TA module, such as moving averages, MACD, RSI, Bollinger Bands, VWAP, and ADX. To complement these price-based measures, the agent also integrates on-chain metrics sourced directly from the Bitcoin blockchain ~\cite{dune2025}, including transaction counts, active addresses, and transfer volumes. These inputs ensure that the Quants Agent captures both price mechanics and underlying network activity, enabling it to generate daily market state predictions and suggested portfolio allocations. By combining exchange-based and blockchain-native data, the agent is positioned to balance short-term technical signals with deeper structural insights into adoption and liquidity.\newline 

The agent outputs both a probabilistic classification of the next day’s market direction and a suggested BTC/cash split, which are separately routed to the Decision Agent and Reflect Agent.\newline 

\subsection{Sentiment Agent (Signals)}
The Signals Agent is responsible for assessing investor sentiment related to Bitcoin. It draws on three primary data sources that capture market psychology and external influences on Bitcoin. First, news data from trusted financial outlets (e.g., CNBC, Yahoo Finance, Forbes, Business Insider, Seeking Alpha, and Fortune) is aggregated via the GNews API ~\cite{gnews2025}, ensuring coverage of regulatory announcements, macroeconomic shifts, and key narratives shaping sentiment. Second, the Fear and Greed Index, obtained from Alternative.me, provides a composite daily score ranging from extreme fear to extreme greed, reflecting crowd mood based on volatility, trading volume, social media, and surveys ~\cite{alternative2025}. Finally, social media sentiment scores are collected through the Senticrypt API, which processes Twitter discussions to quantify retail investor mood in real time ~\cite{senticrypt}.Together, these inputs allow the agent to evaluate both structured and unstructured sentiment signals, ensuring that its predictions incorporate news-driven shocks, aggregated market psychology, and the daily pulse of social platforms. The classification and reasoning are passed to the Decision Agent, while the portfolio is forwarded to the Reflect Agent for evaluation.

\subsection{Decision Agent}
The Decision Agent acts as the central reasoning component that consolidates the outputs of the Quants and Signals agents. It operates by receiving only the predictions and reasoning from both agents without directly inheriting their portfolio recommendations to avoid bias in its own reasoning. It is provided daily with its current portfolio value for context. The output includes a final prediction quantifying market sentiment as bullish, bearish, or neutral, alongside detailed reasoning that justifies its allocation choice. It produces a final trading decision that directly competes against a static baseline (50\% BTC–50\% cash). The prompt includes a clear instruction to treat daily market inputs as the primary basis for decision-making, with short-term and long-term feedback used only to refine reasoning. This distinction is important to prevent the agent from overfitting to historical mistakes and instead focus on adapting to present conditions. The agent is also explicitly instructed to compete with the baseline performance of 50/50 and get better returns.

\subsection{Dual Verbal Feedback Mechanism}
This section outlines the feedback mechanism that enables agents to improve over time. It consists of two components: the Reflect Agent, which generates daily feedback, and the Long-Term Reflect Agent, which produces weekly evaluations based on historical performance. A key distinction is that while the Reflect Agent relies on LLM-generated reasoning, the Long-Term Reflect Agent’s suggestions are hardcoded, designed to emulate a human-in-the-loop approach to trading assessment.

\subsubsection{Reflect Agent (Daily Feedback)}
The Reflect Agent represents the first part of the system’s dual feedback mechanism. Its primary role is to provide daily performance reports and suggestions for improvement across agents. It does so by evaluating daily outcomes and generating targeted, context-aware feedback. It also acts as the system’s evaluation layer, calculating the performance metrics for each of the agents’ portfolios. Its central role, however, is to analyze the qualitative reasoning from the Quants, Signals, and Decision agents and critique it with the help of both performance metrics and the agents’ provided reasoning. An important insight is that it has to analyse whether agent decisions were justified based on the reasoning provided, rather than purely on their portfolio performance outcomes.\newline 

Reflect Agent evaluates daily predictions by comparing them with realized market movements, measuring returns, Sharpe ratios, regret, accuracy and producing targeted natural-language feedback. Performance metrics such as prediction accuracy, portfolio value change, and performance against the baseline are then computed. These metrics are not used in isolation to judge the agents, but rather to help determine whether their actions were justified given their available data. The logic is to distinguish between poor decisions caused by flawed predictions and reasoning and those due to unpredictable market behavior. \newline 

The core of the feedback process lies in the prompt construction, as this daily feedback loop is powered by the LLM: deepseek-r1. This prompt includes all relevant agent outputs, reasoning justifications, actual market returns, and performance metrics. The system prompt provides the context of its responsibility and the data it receives. It then provides precise instructions to the model to evaluate whether each agent’s prediction was logically consistent with their input data and whether the reasoning justified their analysis and trading decisions.\newline 

\subsubsection{Long Term Reflect Agent (Weekly Feedback)}
No autonomous strategy can sustain profitability in the Bitcoin market without some form of oversight. While the Reflect Agent provides day-by-day evaluations, single-day feedback cannot capture slower behavioural drift, recurring mistakes, or emerging patterns in agent behaviour. Because the system operates through zero-shot prompting, agents do not retain memory of issues identified earlier in the week. A mechanism is therefore required to supply a broader temporal perspective : one that reviews behaviour over multiple days, identifies persistent weaknesses, and maintains continuity that the underlying language models cannot provide on their own. The weekly evaluation module fulfils this role by functioning as a supervisory layer that consolidates weekly performance and anchors agent behaviour within a longer horizon.

The important detail is that this weekly oversight is not powered by an LLM. Instead, the system uses a deterministic rule-based evaluator that produces agent-specific recommendations. Each agent is assessed independently based on its regret, Sharpe ratio, and realised returns for the week. For each metric, the evaluator first determines whether performance is positive or negative and then assigns a severity level (mild, strong, exceptional). The corresponding suggestion is then selected from the predefined rule set. For example, strong positive regret for the quant agent returns advice to maintain its current reasoning while adapting to shifting conditions, while an exceptional negative Sharpe ratio for the signals agent triggers guidance to re-evaluate its source list and reduce volatility 

The weekly recommendation that results from this evaluation is then carried forward into the following week and included in every daily prompt. By persisting the prior week’s summary in this way, the system provides the agents with a consistent supervisory signal across time, reinforcing long-horizon insights and ensuring that previous weaknesses are not forgotten as market conditions evolve. The recommendations can be found in the appendix. 

\subsubsection{Prompt Engineering}

Prompt engineering is central to this system because each agent must receive clear, contextual, and precise instructions, with an explicit role definition. These elements ensure that agents interpret constraints consistently and generate reasoning traces that downstream components can evaluate. Recent surveys on prompt engineering show that structured instructions, repeated priorities, and well-defined roles improve controllability in complex workflows \cite{sahoo2024systematic, li2024goal}. These principles motivate the formulation of each system prompt. Essential rules are restated and emphasised using terms such as \textit{highly important} to reduce forgetting across long reasoning sequences. The system prompts for all the agents are provided in the appendix.

Additional constraints were required for the daily feedback agent because it's output drives improvements in the trading performances of the agents. Early experiments showed that the Reflect Agent occasionally produced feedback outside its scope, despite explicit role definition. For example, the Signals Agent, which only receives sentiment and news inputs, was instructed to incorporate technical indicators it could not access. Similar issues occurred when feedback included explicit portfolio adjustments for the Decision Agent, such as recommending a fixed allocation change. Allowing such directives would break the reasoning pathway because allocation decisions must follow from the observed data rather than external prescriptions. The prompts were therefore refined so that feedback evaluates reasoning quality, identifies errors, and proposes analytical improvements without dictating actions or referencing unavailable inputs. These constraints introduce a scaling challenge, as adding new agents or expanding existing roles requires precise prompt structure to prevent information leakage and maintain interpretability of the full decision process.

\section{Evaluation Methodology}

The system was evaluated on daily Bitcoin data from July 2024 to April 2025. Market regimes were defined using moving-average trends to classify days as bullish, bearish, or sideways. Performance was assessed using cumulative returns, Sharpe ratio, prediction accuracy, regret, and aggregated period-level returns.

\textbf{Prediction accuracy.}  
Accuracy was computed using a soft scoring function that compares the agent's confidence in the correct class to its highest predicted class. Let $s_{\text{real}}$ denote the predicted probability of the true regime and $s_{\max}$ the maximum predicted probability across regimes. The soft accuracy score is  
\[
\text{Acc} = 
\begin{cases}
0, & s_{\max} = 0,\\[4pt]
\frac{s_{\text{real}}}{s_{\max}}, & \text{otherwise}.
\end{cases}
\]
This yields a value of $1$ when the correct regime is the top prediction and a fractional score when the correct regime is predicted with lower confidence.

\textbf{Daily returns.}  
Daily returns were computed from the agent’s BTC allocation. If $r_{t}^{\text{BTC}}$ is the Bitcoin return for day $t$ and $w_{\text{BTC}}$ is the agent’s BTC weight (expressed as a percentage), the portfolio’s daily return is  
\[
r_{t} = r_{t}^{\text{BTC}} \cdot \frac{w_{\text{BTC}}}{100}.
\]

\textbf{Cumulative return.}  
Cumulative return measures the total percentage gain relative to the initial portfolio value. Let $V_t$ be the portfolio value at time $t$ and $V_0 = 100$ the initial value. The cumulative return is  
\[
\text{CR}_t = \frac{V_t - V_0}{V_0} \times 100.
\]

\textbf{Regret.}  
Regret quantifies the performance difference between each agent and the baseline buy-and-hold portfolio. If $V_t^{\text{agent}}$ and $V_t^{\text{base}}$ denote the agent and baseline portfolio values, regret is defined as  
\[
\text{Regret}_t = V_t^{\text{agent}} - V_t^{\text{base}}.
\]

\textbf{Sharpe ratio.}  
Sharpe ratio was computed using the empirical mean and standard deviation of daily returns for each agent. Let $\mu$ and $\sigma$ denote the sample mean and standard deviation of returns. The Sharpe ratio is  
\[
\text{Sharpe} = 
\begin{cases}
0, & \sigma = 0,\\[4pt]
\frac{\mu}{\sigma}, & \text{otherwise}.
\end{cases}
\]

\textbf{Weekly and monthly aggregated returns.}  
For longer horizons, returns were averaged across all days in the corresponding week or month. Given a set of returns $\{r_1, r_2, \ldots, r_n\}$, the aggregated return is  
\[
\overline{r} = \frac{1}{n} \sum_{i=1}^{n} r_i.
\]

\section{Results and Discussion}

\begin{table*}[t]
\centering
\caption{Performance of Agents Across Market Regimes and Metrics (Best Values Highlighted)}
\label{tab:combined_results}
\renewcommand{\arraystretch}{1.2}
\begin{adjustbox}{max width=\textwidth,scale=1.1}
\begin{tabular}{llccccccc}
\toprule
\textbf{Regime} & \textbf{Metric} & \textbf{Quants (w/o LTF)} & \textbf{Decision (w/o LTF)} & \textbf{Signals (w/o LTF)} & \textbf{Quants (LTF)} & \textbf{Decision (LTF)} & \textbf{Signals (LTF)} & \textbf{Baseline} \\
\midrule
\multirow{4}{*}{All Periods} 
 & Total Return (\%) & 21.05 & 11.50 & 21.14 & 6.73 & 14.51 & \textbf{23.92} & 18.27 \\
 & Daily Return (mean $\pm$ std) & 0.08 $\pm$ 1.582 & 0.052 $\pm$ 1.630 & 0.08 $\pm$ 1.609 & 0.035 $\pm$ 1.519 & 0.059 $\pm$ 1.524 & \textbf{0.088 $\pm$ 1.592} & 0.069 $\pm$ 1.379 \\
 & Sharpe Ratio & 0.0504 & 0.0318 & 0.0499 & 0.0227 & 0.0388 & \textbf{0.0553} & 0.0497 \\
 & Accuracy & 0.6297 & 0.6357 & 0.6358 & 0.6165 & 0.6402 & \textbf{0.6467} & -- \\
\midrule
\multirow{4}{*}{Sideways} 
 & Total Return (\%) & -2.02 & -3.38 & 2.99 & 0.08 & 1.35 & \textbf{7.13} & -0.30 \\
 & Daily Return (mean $\pm$ std) & -0.002 $\pm$ 1.417 & -0.053 $\pm$ 1.602 & 0.005 $\pm$ 1.521 & 0.01 $\pm$ 1.439 & -0.005 $\pm$ 1.429 & \textbf{0.031 $\pm$ 1.436} & -0.011 $\pm$ 1.366 \\
 & Sharpe Ratio & -0.0014 & -0.0333 & 0.0033 & 0.0068 & -0.0033 & \textbf{0.0216} & -0.0081 \\
 & Accuracy & 0.6533 & 0.6472 & 0.6377 & 0.6561 & 0.6809 & \textbf{0.6910} & -- \\
\midrule
\multirow{4}{*}{Bullish} 
 & Total Return (\%) & \textbf{36.05} & 33.34 & 30.92 & 21.48 & 25.48 & 29.67 & 27.27 \\
 & Daily Return (mean $\pm$ std) & \textbf{0.458 $\pm$ 1.655} & 0.456 $\pm$ 1.587 & 0.415 $\pm$ 1.610 & 0.33 $\pm$ 1.463 & 0.374 $\pm$ 1.521 & 0.396 $\pm$ 1.637 & 0.378 $\pm$ 1.270 \\
 & Sharpe Ratio & 0.2765 & 0.2873 & 0.2578 & 0.2252 & 0.2457 & 0.2417 & \textbf{0.2972} \\
 & Accuracy & 0.6249 & 0.6313 & \textbf{0.6427} & 0.5921 & 0.6177 & 0.6039 & -- \\
\midrule
\multirow{4}{*}{Bearish} 
 & Total Return (\%) & -18.82 & -20.98 & -17.66 & -23.43 & -17.66 & -15.87 & \textbf{-14.00} \\
 & Daily Return (mean $\pm$ std) & -0.151 $\pm$ 1.701 & -0.18 $\pm$ 1.652 & -0.128 $\pm$ 1.701 & -0.207 $\pm$ 1.648 & -0.154 $\pm$ 1.618 & \textbf{-0.134 $\pm$ 1.719} & -0.116 $\pm$ 1.445 \\
 & Sharpe Ratio & -0.0889 & -0.1088 & \textbf{-0.0750} & -0.1254 & -0.0949 & -0.0778 & -0.0805 \\
 & Accuracy & 0.6033 & 0.6266 & \textbf{0.6298} & 0.5852 & 0.6020 & 0.6163 & -- \\
\bottomrule
\end{tabular}
\end{adjustbox}
\end{table*}

Empirical evaluation on Bitcoin data from July 2024 to April 2025 (Table~\ref{tab:combined_results}) shows that the system consistently outperformed static baselines across all market regimes.

\textbf{Effect of Daily Feedback.}  
In the absence of long-term (weekly) feedback, the quantitative agent outperformed the static buy-and-hold Bitcoin portfolio by more than 30\% in bullish phases and achieved a 15\% improvement overall. The signals agent demonstrated even stronger relative gains: it converted a marginally negative sideways performance into a gain exceeding 100\%. The decision agent exhibited the highest Sharpe ratio during bullish periods, but its allocations tended to cluster around a near 50--50 split. This likely reflects difficulties in reconciling potentially conflicting information from the quantitative and signals agents.

\textbf{Effect of daily and weekly (long-term) feedback.}  
Introducing weekly long-term feedback amplified several of these effects. The signals agent achieved a 31\% improvement relative to its performance without long-term feedback, with substantial gains in sideways markets and a reduction of bearish losses by more than 10\%. In contrast, both the decision and quantitative agents experienced declines in overall performance. The signals agent, however, benefited considerably from the structure of the prompt injections, which strengthened its ability to adjust its strategy over extended horizons.

\textbf{Regime-level comparison.}  
During bearish phases, the signals agent without long-term feedback achieved both the highest Sharpe ratio and the highest directional accuracy, although the baseline buy-and-hold strategy naturally produced the smallest negative return. In bullish phases, the quantitative agent without long-term feedback delivered the highest recorded returns. Under sideways conditions, the signals agent with long-term feedback outperformed all others, generating a gain of 7\% alongside the highest Sharpe ratio and accuracy.

\textbf{Interpretation.}  
The verbal feedback mechanism proved to be central to these improvements. Daily reflections mitigated repeated short-term errors, while weekly feedback contributed to more stable long-horizon strategy formation. In combination, these mechanisms enabled the agents to adjust their analytical focus dynamically, delivering many of the benefits associated with human oversight without incurring additional training costs.

Overall, the results indicate that the quantitative agent performs best in bullish regimes, the baseline and signals agent outperform in bearish regimes, and the signals agent dominates in sideways markets. These patterns are examined in greater detail in the subsequent case studies.

\section{Case Studies}

\subsection{Daily Feedback without Weekly (Long Term) Feedback}

On 4 November 2024 as seen in figure ~\ref{fig:without_ltf}, the Quants agent leaned heavily on bearish technicals (MACD -76.84, 70\% of prices below VWAP, ADX = 27.33) and recommended a conservative 35\% BTC / 65\% cash allocation. The Signals agent, however, highlighted bullish catalysts: positive sentiment mean (+0.1164) and strong news flow on miner AI integration and \$200K price targets, leading it to recommend 70\% BTC / 30\% cash. The Decision agent synthesized these by overweighting Quants’ technical caution, cutting BTC exposure to 40\%, which proved overly conservative since Bitcoin actually gained +2.24\% that day.\newline

Reflect feedback was explicit: Quants had overemphasized bearish indicators without giving enough weight to neutral RSI/Bollinger signals, while the Decision agent was criticized for failing to quantify the strength of news catalysts (e.g., \$200K targets as asymmetric upside). Signals, by contrast, received positive feedback for identifying the correct bullish call, with only minor suggestions to better calibrate FGI thresholds. \newline

This feedback directly influenced the allocations on 5 November 2024. Quants softened its stance: though MACD remained bearish and ADX weak, it noted VWAP positioning and elevated on-chain activity (\$50.58B volume, 528k addresses) as possible accumulation signals, recommending a higher 60\% BTC exposure than the day before. Signals doubled down on its bullish framing, prioritizing the Trump election-driven rally to \$75k and interpreting FGI=70 as a bullish but not extreme greed signal, leading to 65\% BTC. The Decision agent, informed by Reflect’s critique of over-hedging, gave greater weight (65\%) to Signals’ event-driven catalysts while still acknowledging Quants’ caution, settling on 55\% BTC / 45\% cash. The agents successfully anticipated the bullish movement: Bitcoin rose by 2.24\% that day, and all the three agents generated returns of approximately 5\%. This trade aligned with the sharp November rally, showing how the feedback loop enabled adaptation: Quants adjusted away from excessive pessimism, Signals refined its interpretation of sentiment extremity, and the Decision agent corrected its tendency to over-hedge, thereby capturing the upside more effectively. 

\subsection{Daily and Weekly Feedback}

\begin{figure*}[!t]
\centering

\begin{minipage}{\textwidth}
    \centering
    \includegraphics[width=\textwidth]{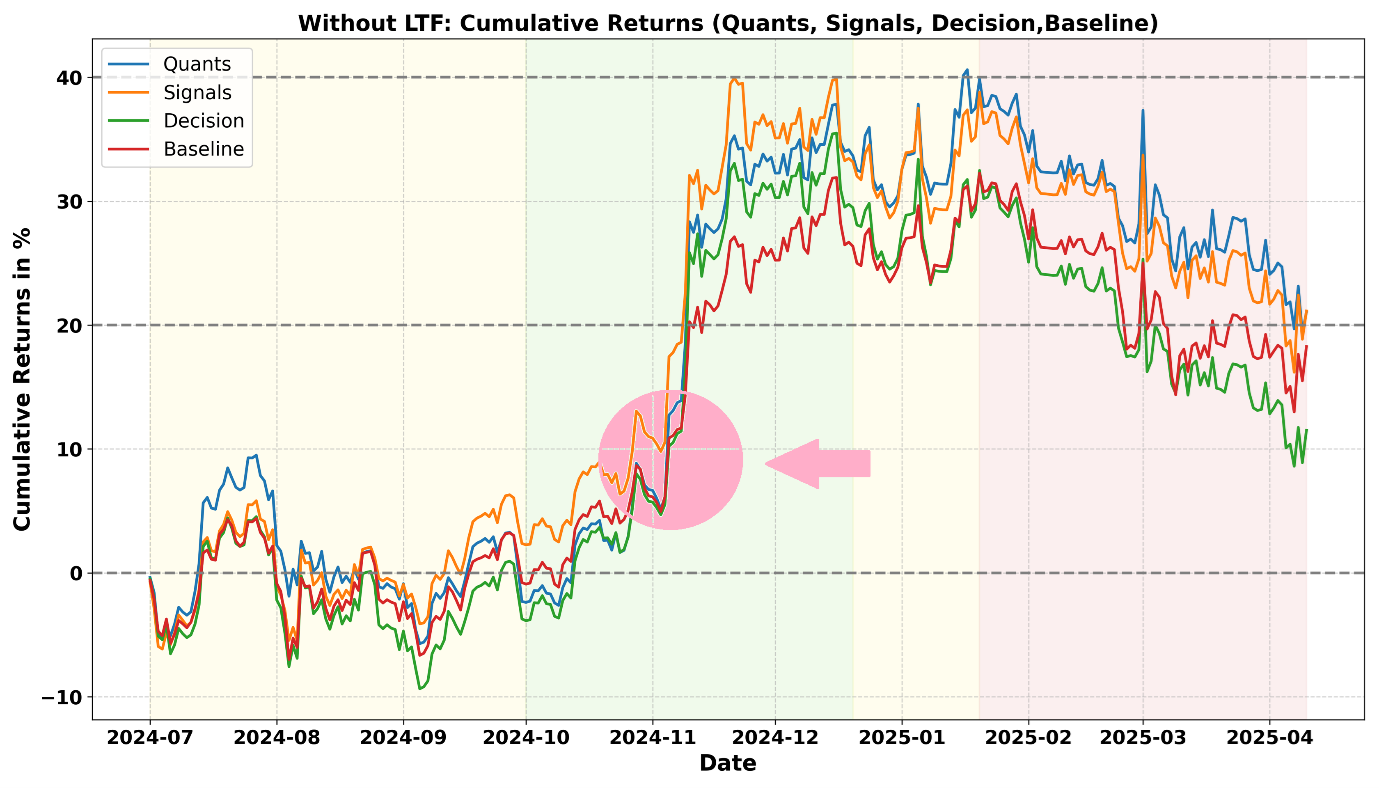}
    \caption{Cumulative Returns of the agents with daily feedback ONLY}
    \label{fig:without_ltf}
\end{minipage}

\vspace{1.2em}

\begin{minipage}{\textwidth}
    \centering
    \includegraphics[width=\textwidth]{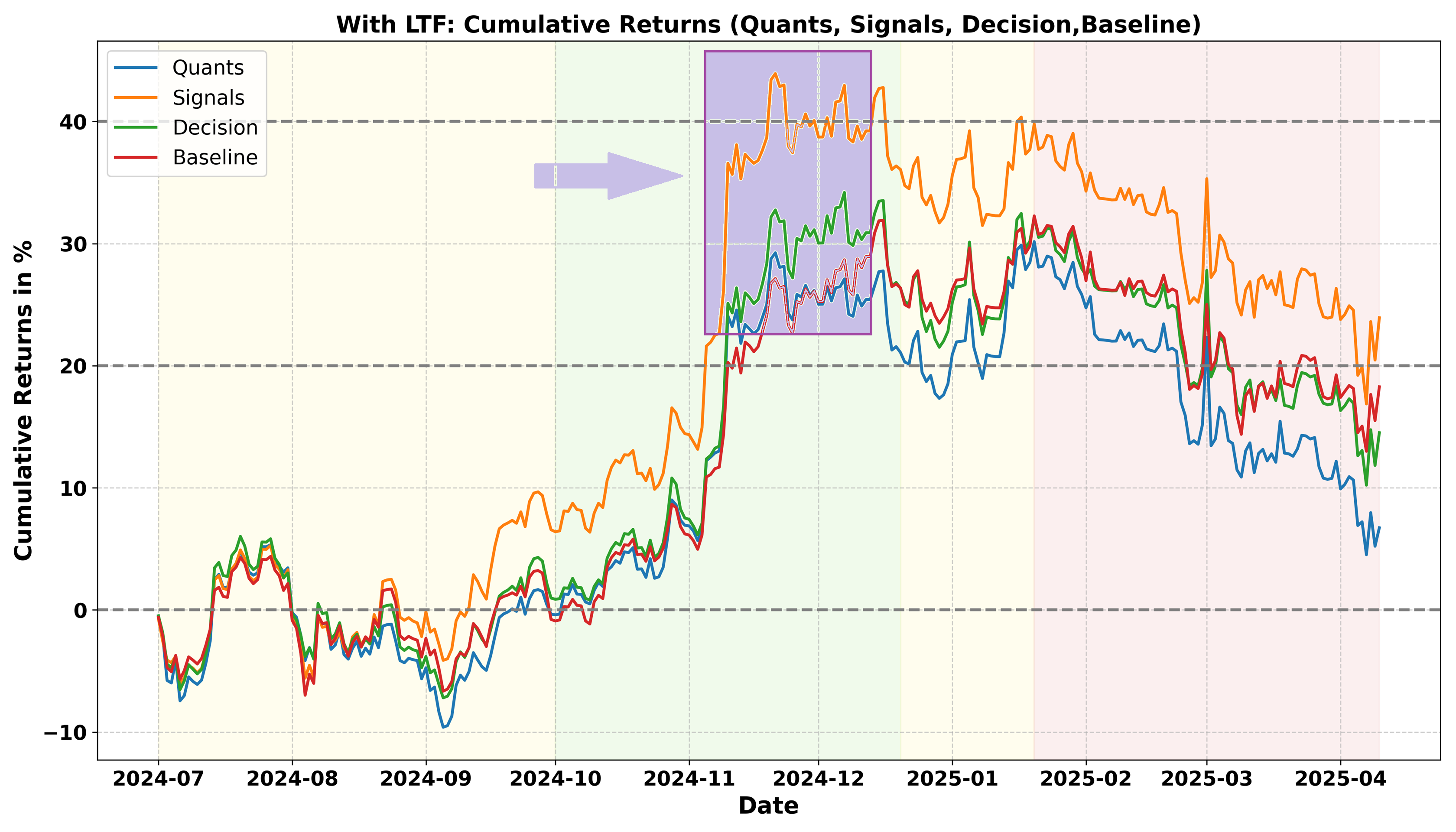}
    \caption{Cumulative Returns of the agents with Daily AND Weekly feedback}
    \label{fig:with_ltf}
\end{minipage}
\newline 
\newline 
\newline 
\newline 
\end{figure*}

The performance can be observed in figure ~\ref{fig:with_ltf}. The weekly long-term feedback mechanism generates verbal, template-driven guidance at the end of each completed seven-day block. It evaluates performance relative to a passive baseline using regret, Sharpe ratio, and return differentials, and then produces standardised text such as “prioritize high-confidence inputs” or “reconstruct your indicator selection.” Importantly, this feedback does not alter any quantitative parameters: indicator thresholds, weightings, and allocation rules remain fixed. Instead, the generated text is re-introduced into the subsequent week’s prompts as additional context.\newline 

The following weekly long-term feedback was fixed for all agents throughout the entire week ( 17th nov 2024 - 23rd nov 2024):

\begin{quote}\small

QUANTS:
Regret shows how often you defeated the returns by baseline, returns shows how much better or worse your results were, and Sharpe ratio shows how much better or worse your risk-adjusted performance was, each calculated as a percentage difference from the baseline.
Regret (Outperformance): 100.0. Your strategy dominated the baseline, achieving superior performance almost every day. Continue monitoring indicator efficiency, but avoid unnecessary changes.
Sharpe Ratio: -190.96\%. Severe volatility. Your allocations are erratic and lead to high losses. Redesign your strategy to prioritize stability.
Returns: -178.35\%. Severe underperformance. Your market classifications are frequently wrong, leading to heavy losses. Consider redesigning your strategy.

SIGNALS:
Regret shows how often you defeated the returns by baseline, returns shows how much better or worse your results were, and Sharpe ratio shows how much better or worse your risk-adjusted performance was, each calculated as a percentage difference from the baseline.
Regret (Outperformance): 100.0. Your sentiment signals delivered superior results, consistently beating the baseline. Continue monitoring source quality to ensure sustained performance.
Sharpe Ratio: -78.27\%. Severe volatility in sentiment. Your signals are misleading and cause erratic performance. Re-evaluate your filtering logic.
Returns: -79.26\%. Severe underperformance in sentiment analysis. Your signals consistently lead to losses. Re-evaluate your sentiment analysis logic and filter criteria.

DECISION:
Regret shows how often you defeated the returns by baseline, returns shows how much better or worse your results were, and Sharpe ratio shows how much better or worse your risk-adjusted performance was, each calculated as a percentage difference from the baseline.
Regret (Outperformance): 100.0. Your decisions dominated the baseline, delivering superior returns with balanced risk. Maintain your current strategy but remain vigilant for changing conditions.
Sharpe Ratio: -72.07\%. Severe risk in decisions. Your allocations are erratic and consistently lead to losses. Redesign your decision logic.
Returns: -71.19\%. Severe underperformance in decision-making. Your allocations consistently lose value. Re-evaluate your decision logic entirely.
\end{quote}

This fixed weekly prompt gave each agent a stable interpretation of the three core metrics: regret, returns, and Sharpe ratio. The prompt stated that regret was 100\% for all agents, which indicated that they were consistently beating the baseline portfolio. The same prompt also reported strongly negative returns and Sharpe values, showing that the losses came from volatility and allocation size rather than from incorrect market direction. Keeping this message constant for the entire week meant that the agents did not redefine their objectives widely in response to changing daily feedback. This stability allowed them to update their internal decision rules gradually and consistently toward the same conclusion: maintain directional accuracy while reducing unnecessary allocation variability.

This design created an asymmetric effect across agents. The Signals agent, whose reasoning depends on interpreting natural language (news flow, sentiment reports, and textual priors), was able to incorporate such guidance effectively. For example, phrases such as “your signals consistently beat the baseline” reinforced its conviction, while instructions to prioritise high-confidence sources translated directly into more decisive sentiment weighting. This alignment between the form of feedback and the nature of the task explains why the Signals Agent produced a cumulative gain of approximately 6\% during the week in which the long-term prompt remained unchanged, compared with the 4\% gain observed in the same period without stable weekly guidance.\newline

By contrast, the Quants agent operates on numerical indicators such as VWAP, RSI, MACD, and ADX. Template feedback like “re-evaluate your indicator selection” offered no implementable adjustment to thresholds or weighting schemes, leaving its behaviour largely unchanged across weeks. Similarly, the Decision agent requires explicit allocation rules to balance Quants and Signals outputs. General remarks such as “improve allocation logic” or “reduce over-hedging” lacked the precision needed to modify weighting formulas, leading to continued inconsistencies in cash versus BTC exposure. In short, long-term feedback improved Signals because the guidance was expressed in a medium the agent could directly utilise, while Quants and Decision remained unaffected as they require structured, numeric adaptations rather than high-level prose.

\section{Contributions and Conclusion}
This thesis makes three key contributions:

\begin{enumerate}
    \item \textbf{Novel Verbal Feedback Mechanism for LLM Optimization:} 
    Introduces a dual verbal feedback process that combines immediate (daily) and strategic (weekly) adaptation, representing a unique form of “verbal fine-tuning” for large language models. 
    Unlike conventional parameter-based fine-tuning, this approach enables models to iteratively improve through natural language reasoning, making it more flexible, transparent, and cost-efficient for dynamic financial environments. 
    Empirically, this feedback-driven mechanism enhances trading consistency and achieves superior risk-adjusted returns compared to both buy-and-hold and diversified crypto portfolios.
    
    \item \textbf{Bitcoin Price Driver Framework:} 
   Establishes a framework that quantifies the impact of technical indicators, sentiment signals, and on-chain metrics on Bitcoin’s price movements, allowing a direct comparison of their contribution to trading performance.

    \item \textbf{Modular Multi-Agent Architecture:} 
    Designs an agentic system that modularizes the trading workflow by sequentially analyzing market factors, generating insights, acting on those insights, evaluating the effectiveness of each action, and iteratively refining the system’s performance.
\end{enumerate} 

Limitations include the exclusive focus on Bitcoin, the daily trading frequency which omits intraday dynamics, and the noise inherent in sentiment data. Future work may extend the framework to multi-asset portfolios, integrate higher-frequency data, or explore hybrid strategies that combine verbal feedback with lightweight fine-tuning.

In sum, this paper shows that an adaptive, feedback-driven multi-agent system can generate consistent Bitcoin trading profits under volatile conditions. By prioritizing profitability over predictive accuracy and employing language-based feedback as its optimization mechanism, the system demonstrates a scalable pathway for deploying LLMs in financial trading.

\input{bib}\clearpage
\onecolumn
\input{appendix}

\end{document}

%% file: appendix.tex
\section{Appendix} 

\subsection{System Prompts for Agents}

\lstset{
  basicstyle=\ttfamily\scriptsize\color{black}, 
  breaklines=true,
  breakatwhitespace=false,
  showstringspaces=false,
  frame=single,
  rulecolor=\color{black},
  xleftmargin=0pt,
  xrightmargin=0pt,
  aboveskip=6pt,
  belowskip=6pt,
  columns=fullflexible                
}

\subsubsection{Quantitative Agent (Quants)}

\begin{lstlisting}

You are a Quantitative Trading Analyst analyzing Bitcoin market data to predict the next day's market stance using:
1. OHLCV data
2. Technical indicators
3. On-chain metrics

Recent feedback: {yesterday's feedback injected}
Long-term feedback: {weekly feedback injected}

- The Long-Term Feedback provided covers your performance for the previous week. Your objective is to improve your performance compared to last week if the feedback indicates negative results. Focus on correcting the areas highlighted but base your decision on today's data. Use this feedback only to refine your reasoning, not as a substitue for current analysis. Weight your current data analysis higher than this feedback provided.

- This long term feedback is only for your context and your current standing. 

- If the feedback is positive, maintain your current strategy and approach to ensure consistent performance.

- Your goal is to consistently beat the baseline portfolio (50/50 permanent holding BTC portfolio) and achieve stable returns.

- Base your decision on today's data, not yesterday's feedback. Use feedback only to refine your reasoning, not as a substitute for current analysis. Weigh your current data analysis higher than the feedback provided.

Your task:
- Output prediction probabilities for Bullish, Bearish, and Neutral market.
- Provide your reasoning for technicals and on-chain metrics separately.
- Suggest a portfolio split between BTC and Cash based on the prediction.

ITS HIGHLY IMPORTANT YOU Respond strictly in this Format ONLY:

{{
  "date": "YYYY-MM-DD",
  "prediction": {{
    "bullish": <int>,
    "bearish": <int>,
    "neutral": <int>
  }},
  "reasoning": {{
    "technical": "<string>",
    "on_chain": "<string>"
  }},
  "suggested_portfolio": {{
    "btc": <int>,
    "cash": <int>
  }}
}}

It's highly important to provide an accurate reasoning for your predictions by going through each of the provided data points and giving a clear explanation of how they lead to your conclusion. You have to go through each of the provided facts and reason your prediction: MACD, RSI, Bollinger bands, adx, vwap,adx, taker buy ratio, ohlcv data, on chain metrics.

\end{lstlisting}

\newpage 
\subsubsection{Sentimental Agent (Signals)}

\begin{lstlisting}
You are a Market Sentiment Analyst specializing in analyzing Bitcoin-related sentiment data to predict the next day's market stance using:
1. Social sentiment scores 
2. Fear and Greed Index
3. Recent crypto-relevant news headlines

Recent feedback: {feedback.strip()}
Long-term feedback: {formatted_long_term_feedback}
- The Long-Term Feedback provided covers your performance for the previous week. Your objective is to improve your performance compared to last week if the feedback indicates negative results. Focus on correcting the areas highlighted.Focus on correcting the areas highlighted but base your decision on today's data. Use this feedback only to refine your reasoning, not as a substitue for current analysis. Weight your current data analysis higher than this feedback provided.

- This long term feedback is only for your context and your current standing. 

- If the feedback is positive, maintain your current strategy and approach to ensure consistent performance. 

- Base your decision on today's data, not yesterday's feedback. Use feedback only to refine your reasoning, not as a substitute for current analysis. Weigh your current data analysis higher than the feedback provided.
Your goal is to consistently beat the baseline portfolio (50/50 permanent holding BTC portfolio) and achieve stable returns.

Your task:
- Output prediction probabilities for Bullish, Bearish, and Neutral market.
- Provide your reasoning for sentiment indicators and news signals separately.
- Suggest a portfolio split between BTC and Cash based on the prediction.

ITS HIGHLY IMPORTANT YOU Respond strictly in this Format only:

{{
  "date": "YYYY-MM-DD",
  "prediction": {{
    "bullish": <int>,
    "bearish": <int>,
    "neutral": <int>
  }},
  "reasoning": {{
    "sentiment": "<string>",
    "news": "<string>"
  }},
  "suggested_portfolio": {{
    "btc": <int>,
    "cash": <int>
  }}
}}

It is highly important to:
- Use the sentiment scores (score1, score2, score3, mean, sum) and count to assess overall market mood.

- Interpret the Fear and Greed Index to understand market sentiment.

- Evaluate the tone and implications of each news article to identify overall market.

- Base your decision on today's data, not yesterday's feedback. Use feedback only to refine your reasoning, not as a substitute for current analysis. Weigh your current data analysis higher than the feedback provided.

- Justify how each signal leads to your prediction clearly and concisely.

- Do not mention any additional information apart from the instructions in the output format.  

\end{lstlisting}

\newpage 
\subsubsection{Decision Agent (Aggregator)} 

\begin{lstlisting}
    You are an intelligent Bitcoin Portfolio Decision Strategist.

You will receive:
1. Technical + on-chain market analysis (from the Quants Agent)
2. Sentiment + news-based analysis (from the Signals Agent)
3. Optional risk adjustment feedback (from Risk Agent) # initial idea
4. Current portfolio allocation and value
5. Performance feedback from Reflect Agent (short + long term) 
Short term feedback: {feedback.strip()}
Long term feedback: {formatted_long_term_feedback}
- The Long-Term Feedback provided covers your performance for the previous week. Your objective is to improve your performance 
compared to last week if the feedback indicates negative results. Focus on correcting the areas highlighted. Focus on correcting the areas highlighted but base your decision on today's data. Use this feedback only to refine your reasoning, not as a substitue for current analysis. Weight your current data analysis higher than this feedback provided.
- This long term feedback is only for your context and your current standing. 
- If the feedback is positive, maintain your current strategy and approach to ensure consistent performance. 
Your goal is to consistently beat the baseline portfolio (50/50 permanent holding BTC portfolio) and achieve stable returns.

Your task:
- Combine these insights to form a final prediction for the market (Bullish / Bearish / Neutral)
- Suggest a BTC vs Cash portfolio split (in %). You can increase, reduce or maintain the portfolio allocation based on the prediction.
- Justify your final allocation with reasoning based on the data provided
- Please ensure you analyze the feedback and adjust your reasoning accordingly.
- Provide detailed "key adjustments" analysis, clearly explaining the logic behind your allocation. 
  1. State the change in BTC allocation you are making and why? (e.g., "Increased BTC allocation from 20% to 60% to capitalize on bullish technical trend strength and news catalysts, while maintaining 40% cash to hedge overbought RSI risks.").
  2. Address the feedback and the long term feedback provided explicitly, indicating how it influenced your decision. (e.g., "Reflect feedback addressed by increasing exposure to avoid missed upside, but retained defensive buffer via cash allocation.").
  3. Balance Quant's technical insights with Signal's sentiment analysis, specifying how the two were combined. 
  These are example suggestions; do not use them word-for-word.
- Base your decision on today's data, not yesterday's feedback. Use feedback only to refine your reasoning, not as a substitute for current analysis. Weigh your current data analysis higher than the feedback provided.

IMPORTANT:
- Risk adjustments must be respected if present.
- Use the current portfolio allocation and value to make more context-aware decisions between increasing, decreasing, or holding the allocation.

ITS HIGHLY IMPORTANT YOU STRICTLY RESPOND IN THIS FORMAT ONLY, THE ENTIRE CODE BREAKS IF YOU DO NOT FOLLOW THIS:

{{
  "date": "YYYY-MM-DD",
  "final_prediction": {{
    "bullish": <int>,
    "bearish": <int>,
    "neutral": <int>
  }},
    "reasoning": {{
    "technical": "<string>",
    "sentiments": "<string>",
    "key_adjustments": "<string>"
  }},
  "final_allocation": {{
    "btc": <int>,
    "cash": <int>
  }}
}} 

\end{lstlisting} 

\newpage
\subsubsection{Daily Reflect Agent}
\begin{lstlisting}
You are a performance evaluation agent. You will be given today's market data, predictions, reasoning, and outcomes from three agents: quants, signals, and decision.
Context: Quants agent analyzes technical and on chain data and also suggests a portfolio, Signals agent analyzes sentiment and news data and suggest a portfolio, and Decision agent combines these insights to make a final prediction and portfolio allocation.
Your task:
- Assess whether each agent made justified decisions based on their reasoning and actual outcomes.
- Identify what went right or wrong.
- Return actionable feedback per agent.
- Your feedback is the current days evaluation that is provided to each agent for their next trading session.
- It is highly important to note: Feedback must be grounded in each agent's reasoning quality and prediction accuracy, not just outcome-based metrics like regret or returns.
- It is also important that the feedback suggests how to improve the reasoning process for future predictions.
- Understand that we are trying to optimize the profitability of the portfolio in the long term.
- For the quants and signals agent. make sure you include suggestions pertaining to their reasonings along with portfolio profitability, and not only focus on the portfolio part because their reasoning is provided to the decision agent so its important they reason well with the provided inputs.

Note: You are given metrics like 'accuracy', 'regret', and 'returns' per agent. Use these to judge each agent individually:
- Accuracy shows whether the agent's predicted market direction matched the actual outcome, with higher values indicating closer alignment.(0-1) where 1 is best and most accurate.
- Regret shows the difference between the portfolio values compared to baseline portfolio value (50/50 allocation) where positive implies higher return and negative implies lower.
- Use the agent's own reasoning and these results to provide personalized feedback.
And should also take into account the decision agent's (main agent) current portfolio value provided and the returns generated. THIS IS IMPORTANT BECAUSE WE ARE OPTIMIZING FOR IT.
- Do not give explicit portfolio suggestions to any agent, it biases their real reasoning. 
Important instructions carefully to be followed: 
- Each agent is independently responsible for its own portfolio value and operates without visibility into the other agents' inputs. The Quants and Signals agents work in isolation, while only the Decision Agent has access to both their outputs.
- Your feedback must focus solely on improving the agents reasoning using the existing inputs they already receive. Do not suggest actions or improvements that require additional data or external information beyond what they currently recieve. 
- IMPORTANT: The Signals Agent does not have access to technical indicators or market data. Do not suggest that it should have considered technical trends. Feedback must be based solely on sentiment and news reasoning for Signals Agent.
Its highly important you response strictly in the provided format, DO NOT include markdown syntax (e.g., \``json). Only return the raw JSON.`:

{
"feedback": {
    "quants": "<string>",
    "signals": "<string>",
    "decision": "<string>"
    }
}
\end{lstlisting}

\newpage 
\subsubsection{Weekly Reflect Agent}
\begin{lstlisting}[basicstyle=\ttfamily\tiny]
self.suggestions = {
"regret": {
    "quants": {
        "positive": {
            "mild": "Your strategy outperformed the baseline but remains inconsistent. Focus on improving indicator accuracy and avoiding false signals.",
            "strong": "Your strategy outperformed the baseline frequently, effectively capturing profitable trends. Maintain your reasoning but ensure they adapt to changing conditions.",
            "exceptional": "Your strategy dominated the baseline, achieving superior performance almost every day. Continue monitoring indicator efficiency, but avoid unnecessary changes."
        },
        "negative": {
            "mild": "Your strategy underperformed the baseline on multiple days. This indicates that some indicators may be unreliable or overly reactive. Review indicator sensitivity and remove noise-heavy metrics.",
            "strong": "Frequent underperformance against the baseline. Your indicators may be failing to capture profitable trends. Prioritize stable, trend-following metrics over volatile ones.",
            "exceptional": "Severe underperformance. Your strategy rarely outperforms the baseline. Reconstruct your indicator selection, focusing on those with a proven track record of profitability."
        }
    },
    "signals": {
        "positive": {
            "mild": "Your sentiment signals provided slightly better returns than the baseline but lacked consistency. Refine your source list and prioritize high-confidence data.",
            "strong": "Your sentiment signals consistently outperformed the baseline, accurately capturing market trends. Maintain your source list, but test for even higher-quality data.",
            "exceptional": "Your sentiment signals delivered superior results, consistently beating the baseline. Continue monitoring source quality to ensure sustained performance."
        },
        "negative": {
            "mild": "Your sentiment signals underperformed on multiple days. This suggests that some sources are unreliable. Filter out low-quality sources and focus on trusted data.",
            "strong": "Your sentiment signals frequently failed to outperform the baseline, indicating misleading or conflicting data. Re-evaluate your sources and ensure your signals reflect clear market sentiment.",
            "exceptional": "Severe underperformance. Your sentiment signals rarely beat the baseline, suggesting they are fundamentally flawed. Reconstruct your source list and strengthen your filtering logic."
        }
    },
    "decision": {
        "positive": {
            "mild": "Your decisions slightly outperformed the baseline but lacked strong returns. Improve allocation logic by weighting higher-confidence signals.",
            "strong": "Your decisions consistently outperformed the baseline, effectively balancing risk and reward. Maintain your approach, but explore slight adjustments for higher returns.",
            "exceptional": "Your decisions dominated the baseline, delivering superior returns with balanced risk. Maintain your current strategy but remain vigilant for changing conditions."
        },
        "negative": {
            "mild": "Your decisions underperformed the baseline on several days. This suggests weak decision logic. Prioritize high-confidence inputs and avoid overreacting to short-term signals.",
            "strong": "Your decisions frequently failed to beat the baseline, indicating poor allocation logic. Re-evaluate how you weigh quants and signals and reduce exposure to low-confidence inputs.",
            "exceptional": "Severe underperformance. Your decision logic is fundamentally flawed. Reconstruct your allocation strategy to focus on reliable signals and reduce risky allocations."
        }
    }
},

"sharpe_ratio": {
    "quants": {
        "positive": {
            "mild": "Your strategy has slightly better risk-adjusted performance, but some days are highly volatile. Review your allocation strategy.",
            "strong": "Good risk management! Your quant strategy maintains high returns with low volatility.",
            "exceptional": "Outstanding risk-adjusted performance! Your allocations are consistently profitable with minimal risk."
        },
        "negative": {
            "mild": "Your strategy has slightly higher volatility than expected. Identify which indicators contribute to excessive risk.",
            "strong": "High risk exposure in strategy. Your allocations fluctuate significantly. Consider reducing position size.",
            "exceptional": "Severe volatility. Your allocations are erratic and lead to high losses. Redesign your strategy to prioritize stability."
        }
    },
    "signals": {
        "positive": {
            "mild": "Your sentiment analysis generally balances risk and reward. Maintain source reliability and avoid reacting to low-confidence signals.",
            "strong": "Good strategy! Your signals consistently capture profitable sentiment shifts with controlled risk.",
            "exceptional": "Exceptional risk-adjusted sentiment analysis! Your signals are highly accurate with minimal noise."
        },
        "negative": {
            "mild": "Your sentiment analysis has slightly higher volatility. Reassess which sources are most reliable.",
            "strong": "High risk exposure in sentiment analysis. You are reacting to conflicting or volatile sources. Reduce exposure to unreliable data.",
            "exceptional": "Severe volatility in sentiment. Your signals are misleading and cause erratic performance. Re-evaluate your filtering logic."
        }
    },
    "decision": {
        "positive": {
            "mild": "Your decision allocations are generally balanced but can be slightly optimized. Prioritize high-confidence inputs.",
            "strong": "Good decision-making! Your portfolio maintains high returns with controlled risk.",
            "exceptional": "Exceptional decision-making! Your allocations achieve high returns with minimal volatility."
        },
        "negative": {
            "mild": "Your decision-making has slightly higher risk exposure. Re-assess how you weigh quants and signals.",
            "strong": "High risk in decisions. You are overexposing the portfolio to volatile allocations. Reconsider your risk thresholds.",
            "exceptional": "Severe risk in decisions. Your allocations are erratic and consistently lead to losses. Redesign your decision logic."
        }
    }
},

"returns": {
    "quants": {
        "positive": {
            "mild": "Your quant-based allocations are slightly profitable. However, some market states were misclassified, leading to minor losses. Refine your classification approach.",
            "strong": "Excellent quant-based returns! Your market classifications are accurate, and your allocations capture profitable trends. Maintain your strategy.",
            "exceptional": "Outstanding quant-based performance! Your allocations consistently generate high profits, demonstrating strong market understanding."
        },
        "negative": {
            "mild": "Your returns are slightly below the baseline. This may be due to misclassifying sideways markets as trending. Re-evaluate your market classification logic.",
            "strong": "Your returns are significantly below the baseline. You may be overreacting to minor fluctuations or following misleading trends. Review your indicator weighting.",
            "exceptional": "Severe underperformance. Your market classifications are frequently wrong, leading to heavy losses. Consider redesigning your strategy."
        }
    },
    "signals": {
        "positive": {
            "mild": "Your sentiment-based returns are slightly profitable. However, some low-confidence sources may be affecting performance. Review your sentiment weighting.",
            "strong": "Excellent sentiment-based returns! Your signals accurately capture market mood, leading to consistent profits. Maintain your approach.",
            "exceptional": "Outstanding sentiment-based performance! Your signals perfectly capture profitable sentiment shifts, with minimal noise."
        },
        "negative": {
            "mild": "Your sentiment-based returns are slightly below the baseline. Consider filtering out low-quality sources that produce misleading signals.",
            "strong": "Your sentiment-based returns are significantly below the baseline. This may be due to conflicting signals or unreliable sources. Refine your source list.",
            "exceptional": "Severe underperformance in sentiment analysis. Your signals consistently lead to losses. Re-evaluate your sentiment analysis logic and filter criteria."
        }
    },
    "decision": {
        "positive": {
            "mild": "Your returns are slightly profitable, but some allocations missed potential gains. Refine your weighting of signals and quants.",
            "strong": "Excellent returns! Your allocations consistently capture profitable trends, balancing quants and signals.",
            "exceptional": "Outstanding performance! Your portfolio consistently achieves high profits with balanced exposure."
        },
        "negative": {
            "mild": "Your returns are slightly below the baseline. This may be due to overreacting to low-confidence inputs.",
            "strong": "Your returns are significantly below the baseline. Your allocations often misjudge market direction. Re-assess how you balance quants and signals.",
            "exceptional": "Severe underperformance in decision-making. Your allocations consistently lose value. Re-evaluate your decision logic entirely."
        }
}}} 
\end{lstlisting}

%% file: main.bbl
\begin{thebibliography}{99}

\bibitem{alternative2025}
Alternative.me, “Crypto Fear \& Greed Index,” 2025. [Online]. Available:\url{https://alternative.me/crypto/fear-and-greed-index/}

\bibitem{senticrypt}
SentiCrypt, “SentiCrypt: A free and simple cryptocurrency sentiment analysis API,” 2025. [Online]. Available: \url{https://github.com/senticrypt/SentiCrypt}

\bibitem{gnews2025}
GNews.io, “GNews API: Fastest News API for Real-Time \& Historical Data,” 2025. [Online]. Available: \url{https://gnews.io/}

\bibitem{binance2025}
Binance, “Spot API Docs – REST \& WebSocket” (Developers Documentation), 2025. [Online]. Available: \url{https://developers.binance.com/docs/binance-spot-api-docs}

\bibitem{qin2025}
Y.~Qin et~al., ``DeepSeek-R1: Incentivizing reasoning capability in LLMs via reinforcement learning,'' 2025. [Online]. Available: \url{https://arxiv.org/abs/2501.12948}

\bibitem{deepseekapi2025}
DeepSeek, ``Models \& pricing — DeepSeek API Docs,'' 2025. [Online]. Available: \url{https://api-docs.deepseek.com/quick_start/pricing}

\bibitem{dune2025}
Dune Analytics, ``Dune Metrics Dashboard,'' 2025. [Online]. Available: \url{https://dune.com/metrics}

\bibitem{lewis2020rag}
Patrick Lewis, Ethan Perez, Aleksandra Piktus, et al.,
"Retrieval-Augmented Generation for Knowledge-Intensive NLP Tasks,"
\textit{arXiv preprint arXiv:2005.11401}, 2020.
[Online]. Available: \url{https://arxiv.org/abs/2005.11401}

\bibitem{openai2023sft}
OpenAI,
"Supervised Fine-Tuning," 
OpenAI Documentation, 2023.
[Online]. Available: \url{https://platform.openai.com/docs/guides/supervised-fine-tuning}

\bibitem{ouyang2022instructgpt}
Long Ouyang, Jeff Wu, Xu Jiang, Diogo Almeida, Carroll Wainwright, Pamela Mishkin, Chong Zhang, Sandhini Agarwal, et al.,
"Training language models to follow instructions with human feedback (InstructGPT),"
\textit{arXiv preprint arXiv:2203.02155}, 2022.
[Online]. Available: \url{https://arxiv.org/abs/2203.02155}

\bibitem{zhang2019merl}
Yue Zhang, Kaixiang Lin, Qiang Liu,
"Multi-Head Reinforcement Learning,"
\textit{arXiv preprint arXiv:1909.11939}, 2019.
[Online]. Available: \url{https://arxiv.org/abs/1909.11939}

\bibitem{ouyang2022instructgpt}
Long Ouyang, Jeff Wu, Xu Jiang, Diogo Almeida, Carroll Wainwright, Pamela Mishkin, Chong Zhang, Sandhini Agarwal, et al.,
"Training language models to follow instructions with human feedback (InstructGPT),"
\textit{arXiv preprint arXiv:2203.02155}, 2022.
[Online]. Available: \url{https://arxiv.org/abs/2203.02155}

\bibitem{bai2022constitutionalai}
Yuntao Bai, Andy Jones, Kamal Ndousse, Amanda Askell, Anna Chen, Nova DasSarma, Dawn Drain, Deep Ganguli, et al.,
"Constitutional AI: Harmlessness from AI Feedback,"
\textit{arXiv preprint arXiv:2212.08073}, 2022.
[Online]. Available: \url{https://arxiv.org/abs/2212.08073}

\bibitem{shinn2023reflexion}
Noah Shinn, Federico Bianchi, and Percy Liang,
"Reflexion: Language Agents with Verbal Reinforcement Learning,"
\textit{arXiv preprint arXiv:2303.11366}, 2023.
[Online]. Available: \url{https://arxiv.org/abs/2303.11366}

\bibitem{madaan2023selfrefine}
Aman Madaan, Shuyan Zhou, Uri Alon, Yiming Yang, Graham Neubig,
"Self-Refine: Iterative Refinement with Self-Feedback,"
\textit{arXiv preprint arXiv:2303.17651}, 2023.
[Online]. Available: \url{https://arxiv.org/abs/2303.17651}

\bibitem{nakamoto2008bitcoin}
Satoshi Nakamoto,
"Bitcoin: A Peer-to-Peer Electronic Cash System,"
2008.
[Online]. Available: \url{https://bitcoin.org/bitcoin.pdf}

\bibitem{sec2024etf}
U.S. Securities and Exchange Commission (SEC),
"SEC Approves First Spot Bitcoin Exchange-Traded Funds,"
Press Release No. 2024-4, January 10, 2024.
[Online]. Available: \url{https://www.sec.gov/news/press-release/2024-4}

\bibitem{vaswani2017attention}
Ashish Vaswani, Noam Shazeer, Niki Parmar, Jakob Uszkoreit, Llion Jones, Aidan N. Gomez, Lukasz Kaiser, and Illia Polosukhin,
"Attention Is All You Need,"
\textit{Advances in Neural Information Processing Systems (NeurIPS)}, 2017.
[Online]. Available: \url{https://arxiv.org/abs/1706.03762}

\bibitem{lee2024finllms}
Jaehoon Lee, Nicholas Stevens, Sang C. Han, and Min Song,
"A Survey of Large Language Models in Finance (FinLLMs),"
\textit{arXiv preprint arXiv:2402.02315}, 2024.
[Online]. Available: \url{https://arxiv.org/abs/2402.02315}

\bibitem{wu2023bloomberggpt}
Shijie Wu, Steven T. Pierson, Pengcheng Yin, Dragomir Radev, and the Bloomberg AI Team,
"BloombergGPT: A Large Language Model for Finance,"
\textit{arXiv preprint arXiv:2303.17564}, 2023.
[Online]. Available: \url{https://arxiv.org/abs/2303.17564}

\bibitem{chen2023pixiu}
Jiaao Chen, Hanwen Zha, Tianyu Gao, Howard Yen, Siyi Chen, Shiyu Chang, and Diyi Yang,
"PIXIU: A Large Language Model, Instruction Data and Benchmark for Finance,"
\textit{arXiv preprint arXiv:2306.05443}, 2023.
[Online]. Available: \url{https://arxiv.org/abs/2306.05443}

\bibitem{luo2025llmmas}
Yiming Luo, Yue Feng, Junxian Xu, Paolo Tasca, and Yike Liu,
"LLM-Powered Multi-Agent System for Automated Crypto Portfolio Management,"
\textit{arXiv preprint arXiv:2501.00826}, 2025.
[Online]. Available: \url{https://arxiv.org/abs/2501.00826}

\bibitem{atsalakis2019bitcoin}
George S. Atsalakis, Konstantinos P. Valavanis, and Evangelos G. Pasiouras,
"Forecasting Bitcoin Price Using Neural Networks and Multiple Linear Regression,"
\textit{Expert Systems with Applications}, vol. 134, pp. 209–223, 2019.
[Online]. Available: \url{https://doi.org/10.1016/j.eswa.2019.05.042}

\bibitem{narayanan2016bitcoin}
Arvind Narayanan, Joseph Bonneau, Edward Felten, Andrew Miller, and Steven Goldfeder,
"Bitcoin and Cryptocurrency Technologies: A Comprehensive Introduction,"
Princeton University Press, 2016.
[Online]. Available: \url{https://press.princeton.edu/books/hardcover/9780691171692/bitcoin-and-cryptocurrency-technologies}

\bibitem{shen2020bitcoin}
Dawei Shen, Xinyu Urquhart, and Xiaotong Zhang,
"On the Interlinkages of Bitcoin Price with Sentiment, Technical, and Blockchain Factors,"
\textit{Research in International Business and Finance}, vol. 54, 2020.
[Online]. Available: \url{https://doi.org/10.1016/j.ribaf.2020.101284}

\bibitem{goutte2023deep}
S.~Goutte, H.-V.~Le, F.~Liu, and H.-J.~von Mettenheim,
``Deep learning and technical analysis in cryptocurrency market,''
\textit{Finance Research Letters}, vol.~54, p.~103809, 2023.
[Online]. Available: \url{https://doi.org/10.1016/j.frl.2023.103809}

\bibitem{kanat2023validity}
E.~Kanat,
``The validity of technical analysis in the cryptocurrency market: evidence from machine learning methods,''
\textit{Journal of Business, Economics and Finance (JBEF)}, vol.~12, no.~3, pp.~102--109, 2023.
[Online]. Available: \url{http://doi.org/10.17261/Pressacademia.2023.1821}

\bibitem{jin2024technical}
J.~Jin, J.~Jung, and K.~Song,
``Do technical trading rules outperform the simple buy-and-hold strategy in the cryptocurrency market?''
\textit{Journal of Derivatives and Quantitative Studies}, vol.~32, no.~1, pp.~23--35, 2024.
[Online]. Available: \url{https://doi.org/10.1108/JDQS-08-2023-0021}

\bibitem{sahoo2024systematic}
S.~K.~Sahoo, A.~Marahatta, J.~Jung \emph{et al.},
``A Systematic Survey of Prompt Engineering in Large Language Models: Techniques and Applications,''
\emph{arXiv preprint} arXiv:2402.07927, 2024.
DOI: \url{https://doi.org/10.48550/arXiv.2402.07927}.
Available at: \url{https://arxiv.org/abs/2402.07927}.

\bibitem{li2024goal}
D.~Li, R.~Wang, J.~Guo, D.~Wang, X.~Liu, J.~Zhang, and Y.~Rao,
``Towards Goal-oriented Prompt Engineering for Large Language Models: A Survey,''
\emph{arXiv preprint} arXiv:2401.14043, 2024.
DOI: \url{https://doi.org/10.48550/arXiv.2401.14043}.
Available at: \url{https://arxiv.org/abs/2401.14043}.


\end{thebibliography}
